\documentclass[12pt]{article}
 \pdfoutput=1
\usepackage{amssymb}
\usepackage{amsmath}
\usepackage{amstext}
\usepackage{graphicx,epsfig}
\usepackage{epsfig}
\usepackage{verbatim} 
\usepackage{fancyhdr}
\usepackage{fancybox}
\usepackage{color}
\usepackage{ulem,bbold}
\usepackage{enumitem}
\usepackage{subfigure}
\usepackage{bbm}
\usepackage{parskip}

\usepackage{latexsym,amsfonts,amsmath,amssymb,theorem,dsfont,wasysym}
\usepackage{amssymb}
\usepackage{amsmath}
\usepackage{amstext}
\usepackage{graphicx,epsfig}
\usepackage{epsfig}
\usepackage{verbatim} 
\usepackage{fancyhdr}
\usepackage{fancybox}
\usepackage{color}
\usepackage{ulem,bbold}
\usepackage{enumitem}
\usepackage{subfigure}
\usepackage{bbm}
\usepackage{parskip}
\usepackage[numbers,sort&compress]{natbib}

\newcommand{\Comment}[1]{{}}
\definecolor{MyDarkBlue}{rgb}{0.15,0.15,0.45}
\usepackage[linktocpage=true]{hyperref}
\hypersetup{
colorlinks=true,
citecolor=MyDarkBlue,
linkcolor=MyDarkBlue,
urlcolor=MyDarkBlue,
}

\newcommand{\bea}{\begin{eqnarray}}  
\newcommand{\eea}{\end{eqnarray}}
\newcommand{\nn}{\nonumber}

\def\gt{\rightarrow}

\def\be#1\ee{\begin{align}#1\end{align}}
\def\del{\partial}

\def\({\left(}
\def\){\right)}
\def\[{\left[}
\def\]{\right]}

\def\half{\frac{1}{2}}

\def\ddt{\frac{\mathrm{d}}{\mathrm{d} t}}

\def\dy{\mathrm{d} y \,}

\def\dcx{\mathrm{d}^3 x \,}

\def\dfx{\mathrm{d}^4 x \,}

\def\I{\mathcal{I}}

\def\R{\mathcal{R}}

\def\pb{\bar{\phi}}

\def\a{\alpha}
\def\b{\beta}
\def\g{\gamma}
\def\d{\delta}

\def\l{\lambda}

\def\s{\sigma}

\def\f{\varphi}

\def\D{\Delta}

\def\pb{\bar{\phi}}

\def\dy{\mathrm{d} y \,}

\def\nn{\nonumber}

\def\nl{\nonumber \\ &}
\def\nc{\, , \qquad}

\begin{document}

\begin{center}
{\Large \bf{DBI Realizations of the Pseudo-Conformal Universe and Galilean Genesis Scenarios}}
\end{center} 
 \vspace{1truecm}
\thispagestyle{empty} \centerline{
{\large  {Kurt Hinterbichler${}^{a,b}$, Austin Joyce${}^{b}$, Justin Khoury${}^{b}$ and Godfrey E. J. Miller${}^{b}$}}
}

\vspace{1cm}

\centerline{{\it ${}^a$ 
Perimeter Institute for Theoretical Physics,}}
 \centerline{{\it 31 Caroline St. N, Waterloo, Ontario, Canada, N2L 2Y5}} 
 
 \vspace{1cm}

\centerline{{\it ${}^b$ 
 Center for Particle Cosmology, Department of Physics \& Astronomy, University of Pennsylvania,}}
 \centerline{{\it  209 South 33rd Street, Philadelphia, PA 19104}}

 \vspace{1cm}

\begin{abstract}
The pseudo-conformal universe is an alternative to inflation in which the early universe is described by a conformal field theory on approximately flat space-time.  The fields develop time-dependent expectation values, spontaneously breaking the conformal symmetries to a de Sitter subalgebra, and fields of conformal weight zero acquire a scale invariant spectrum of perturbations.  In this paper, we show that the pseudo-conformal scenario can be naturally realized within theories that would ordinarily be of interest for DBI inflation, such as the world-volume theory of a probe brane in an AdS bulk space-time. In this approach, the weight zero spectator field can be associated with a geometric flat direction in the bulk, and its scale invariance is protected by a shift symmetry.
\end{abstract}


\newpage
\setcounter{page}{1}
\tableofcontents

\section{Introduction}

While the observational evidence for primordial adiabatic density perturbations with nearly scale invariant and gaussian statistics is consistent with the predictions of the simplest models of inflation~\cite{Starobinsky:1979ty, Guth:1980zm, Albrecht:1982wi, Linde:1981mu}, it is good scientific practice to seek alternative explanations for the data. Over the years, this has motivated cosmologists to propose alternatives such as, for example, pre-big bang cosmology \cite{Gasperini:1992em, Gasperini:2002bn, Gasperini:2007vw}, string gas cosmology \cite{Brandenberger:1988aj, Nayeri:2005ck, Brandenberger:2006xi, Brandenberger:2006vv, Brandenberger:2006pr, Battefeld:2005av}, the ekpyrotic scenario~\cite{Khoury:2001wf, Donagi:2001fs, Khoury:2001bz, Khoury:2001zk, Lyth:2001pf, Brandenberger:2001bs, Steinhardt:2001st, Notari:2002yc, Finelli:2002we, Tsujikawa:2002qc, Gratton:2003pe, Tolley:2003nx, Craps:2003ai, Khoury:2003vb, Khoury:2003rt, Khoury:2004xi, Creminelli:2004jg, Lehners:2007ac, Buchbinder:2007ad, Buchbinder:2007tw, Buchbinder:2007at, Creminelli:2007aq, Koyama:2007mg, Koyama:2007ag, Lehners:2007wc, Lehners:2008my, Lehners:2009qu, Khoury:2009my, Khoury:2011ii, Joyce:2011ta}, and superluminal scenarios~\cite{ArmendarizPicon:2003ht,ArmendarizPicon:2006if,Piao:2006ja,Magueijo:2008pm,Magueijo:2008sx,Piao:2008ip,Bessada:2009ns,Bessada:2012zx}.

Assuming a single scalar degree of freedom coupled minimally to Einstein gravity, the combined requirements of a spectrum of curvature perturbations that is scale invariant and gaussian over many decades of modes,
a dynamical attractor background, and subluminal propagation leads one to inflation~\cite{Khoury:2010gw,Joyce:2011kh, Baumann:2011dt,Geshnizjani:2011dk}. Therefore, alternative mechanisms which generate perturbations while remaining weakly-coupled must either rely on an instability, as in the contracting matter-dominated scenario~\cite{Wands:1998yp, Finelli:2001sr}, rely on superluminality, as in tachyacoustic cosmology~\cite{Bessada:2009ns}, and/or must involve additional degrees of freedom, as in the New Ekpyrotic scenario~\cite{Buchbinder:2007ad,Lehners:2007ac,Creminelli:2007aq}. 

The {\it pseudo-conformal universe}~\cite{Hinterbichler:2011qk,Hinterbichler:2012mv} is a general framework for describing early universe scenarios
that rely on the spontaneous symmetry breaking of the conformal algebra down to its de Sitter subalgebra
\be
\label{symmbreak}
so(4,2) \longrightarrow so(4,1)
\ee
to generate scale invariant perturbations.\footnote{This is in contrast to the  inflationary universe, which relies on the symmetry breaking pattern $so(4,1)\to so(3)\times\mathbb{R}^3\times{\rm shift}$, to generate scale-invariant perturbations.} These include the quartic $U(1)$-invariant model~\cite{Craps:2007ch,Rubakov:2009np,Osipov:2010ee,Libanov:2010nk,Libanov:2010ci,Libanov:2011hh,Libanov:2011bk} and Galilean Genesis scenarios~\cite{Creminelli:2010ba,LevasseurPerreault:2011mw,Liu:2011ns,Qiu:2011cy,Wang:2012bq,Liu:2012ww,Creminelli:2012my}. In its most general form, the scenario postulates that the early universe is described by a conformal field theory containing conformal scalar fields (which may or may not be fundamental)
\be
\phi_I \nc I = 1 \ldots N \, ,
\ee
each with its own conformal weight $\D_I$. The theory must be chosen so that the fields $\phi_I$ develop time-dependent expectation values
\be
\pb_I = \frac{\a_I}{(-t)^{\D_I}} \, ,
\label{expectationvalues}
\ee
which spontaneously break $so(4,2)$ conformal symmetry down to an $so(4,1)$ de Sitter sub-algebra.\footnote{Pseudo-conformal symmetry breaking requires at least one $\pb_I$ with non-trivial time dependence, {\it i.e.}, for which $\D_I \neq 0$ and $\a_I \neq 0$.} The spontaneous symmetry breaking pattern~(\ref{symmbreak}) is the characteristic signature of a pseudo-conformal scenario.  The residual de Sitter symmetry drives spectator fields of conformal weight zero to acquire scale-invariant perturbations, exactly as if they lived on an inflating background.\footnote{Unitarity bounds~\cite{Mack:1975je} forbid weight-0 perturbations, but assume a stable conformally invariant vacuum.   We do not require such a vacuum, we only need the symmetry-breaking background~(\ref{phibackgenmany}).  The conformal vacuum $\bar\phi_I = 0$ can be unstable or not exist.  In Rubakov's quartic scenario~\cite{Rubakov:2009np}, for instance, the weight-0 mode is the angular component of a complex scalar field, which is of course ill-defined around the unstable trivial background.} Insofar as scale invariance is achieved in entropy perturbations which must later be made adiabatic, the pseudo-conformal mechanism is analogous to the curvaton mechanism~\cite{Lyth:2001nq,Lyth:2002my} or the New Ekpyrotic scenario~\cite{Buchbinder:2007at}. (See~\cite{Wang:2012bq} for a recent discussion of entropy to adiabatic conversion in the pseudo-conformal and Galilean Genesis scenarios.) The standard pseudo-conformal scenario with linear realization of the conformal algebra is reviewed in Sec.~\ref{linearpseudoconformal}.

In this paper, we achieve the characteristic $so(4,2) \to so(4,1)$ pseudo-conformal symmetry breaking starting from a non-linear realization of the conformal algebra. The simplest example is the Dirac--Born--Infeld (DBI) action, obtained from the world-volume of a flat brane probing an ${\rm AdS}_5$ bulk, with an additional tadpole term governed by a dimensionless parameter $\lambda$,
\be
S_{\rm DBI} = \int \dfx \phi^4 \( 1 + \frac{\l}{4} - \sqrt{1+\frac{(\del \phi)^2}{\phi^4}} \) .
\label{diracborninfeld}
\ee
Geometrically, the field $\phi$ represents the transverse displacement of the brane into the radial direction of the bulk AdS. This action inherits the bulk isometries, which act non-linearly as the conformal algebra in the world-volume theory with the field $\phi$ transforming as a weight one field. We will see in Sec.~\ref{dbisymmetrybreaking} that when $\lambda>0$ this action admits a scaling solution $\phi\sim 1/ t$, which breaks the symmetry in the desired way \eqref{symmbreak}, giving a non-linear version of the rolling scalar scenario studied in \cite{Hinterbichler:2011qk}. We show that perturbations around this solution have a sound speed strictly less than one, in contrast with the linear scenario where the sound speed is always equal to one.

One motivation to study this DBI incarnation of the pseudo-conformal universe is that the DBI action and its multi-field generalizations have been used to great effect for inflationary model building~\cite{Alishahiha:2004eh,Silverstein:2003hf,Chen:2005ad,Lidsey:2007gq,Easson:2007dh,Huang:2007hh,Bean:2007eh,Langlois:2009ej,Mizuno:2009mv}. In particular, the search for realistic inflationary models within string theory has often focused on brane inflation scenarios where the inflaton is interpreted as the modulus of a 3-brane probing some AdS-like region of a warped geometry~\cite{Kachru:2003sx}.  The DBI action \eqref{diracborninfeld} and its extensions describe the world-volume dynamics of this process.  If a new substitute for inflation, such as the pseudo-conformal mechanism, can be realized naturally in these setups, it will provide another route by which these scenarios could make contact with the real world.

Another motivation is to realize novel DBI extensions of the Galilean Genesis scenario~\cite{Creminelli:2010ba} with strictly subluminal sound speed.
To do so requires including higher-derivative terms to the lowest-order DBI action~(\ref{diracborninfeld}). Those that give second order
equations of motion are the conformal DBI galileon terms~\cite{deRham:2010eu,Goon:2012dy,Goon:2011uw,Goon:2011qf,Trodden:2011xh,Goon:2011xf,Goon:2012mu}, which themselves have proven useful for inflationary model
building~\cite{RenauxPetel:2011uk,Burrage:2010cu}. We focus here on the cubic conformal DBI galileon term, whose ``non-relativistic" cubic galileon counterpart is the
basis of the Galilean Genesis scenario~\cite{Creminelli:2010ba}.\footnote{The ordinary cubic conformal galileon has been of interest recently in connection to the $a$-theorem --- the coefficient of the cubic conformal galileon term in the effective action for the dilaton encodes the difference in the $a$-anomaly between UV and IR ends of an RG flow in 4D~\cite{Komargodski:2011vj,Komargodski:2011xv}. (The cubic conformal galileon is the only conformal galileon in four dimensions which is a Wess--Zumino term \cite{Goon:2012dy}.)}
One of the drawbacks of the original Galilean
Genesis scenario is superluminality --- linear perturbations propagate exactly at the speed of light around the $1/t$ solution, and perturbations
can tip the sound speed over the edge. In Sec.~\ref{conformalDBIgalileon} we show that Galilean Genesis can be realized with the DBI conformal galileons, that it shares
many of the same features as the non-DBI version, but it has the bonus that perturbations are strictly subluminal. (Alternatively, subluminal Genesis can be realized by 
explicitly breaking part of the conformal symmetries~\cite{Creminelli:2012my}.)

To generate scale invariant entropy perturbations, we need a weight-0 field.  One natural way this can come about is if the bulk space has additional isometry directions besides the AdS, as is often the case in brane inflation scenarios.  As a simplest example, we study in Sec.~\ref{dbiscaleinvariance} an AdS$_5 \times S^1$, bulk.  In this model, the field parameterizing the displacement of the brane into the $S^1$ is an angular weight-0 field which is protected by a shift invariance inherited from the isometry of the circle, and it acquires a scale invariant spectrum of perturbations.

We then study in Sec.~\ref{genquadDBI} the general scenario, using the non-linear DBI symmetries and the breaking pattern to derive the general form of the quadratic fluctuations, showing that the important features of the mechanism --- such as speed of fluctuations and spectrum of perturbations --- are insensitive to the specific realization and depend only on the symmetries. Finally, we discuss the coupling to Einstein gravity in Sec.~\ref{gravity} and verify that cosmological evolution is negligible at early times during pseudo-conformal symmetry breaking. We conclude with a brief discussion of future avenues to pursue in Sec.~\ref{conclu}. In Appendix \ref{cosetrelations}, we comment on the relation between the Galilean Genesis scenario and the nonlinear starting point which is the subject of this work.

\section{Review of the Linear Pseudo-Conformal Scenario}
\label{linearpseudoconformal}

We start by reviewing the general pseudo-conformal scenario as discussed in \cite{Hinterbichler:2011qk}.  We refer to this as the linear pseudo-conformal scenario (since the conformal symmetry is realized linearly on the full fields) to distinguish from the DBI pseudo-conformal scenario we discuss in this paper (where the conformal symmetry is realized non-linearly on the full fields).

In a pseudo-conformal scenario, the early universe is dominated by a conformal field theory containing elementary or composite conformal scalars $\phi_I$, indexed by $I = 1,\ldots, N$, with conformal weights $\Delta_I$.  There are fifteen symmetries which act on $\phi_I$ as
\bea 
\nonumber
\delta_{P_\mu}\phi_I &=&-\partial_\mu\phi_I \,, \qquad  \,\,\,\,\,\,\,\,\,\,\,\,\,\,\,\,\,\,\,\,\,\,\,\, \delta_{J^{\mu\nu}}\phi_I  = (x^\mu\partial^\nu-x^\nu\partial^\mu)\phi_I \,,\\
\delta_D\phi_I &=&- (\Delta_I+ x^\mu\partial_\mu) \phi_I \,, \qquad  \delta_{K_\mu}\phi_I  = \(-2\Delta_Ix_\mu -2x_\mu x^\nu\partial_\nu +x^2\partial_\mu\)\phi_I \,.
\label{delphiconfmany}
\eea
Here $P_\mu$ and $J_{\mu\nu}$ generate the usual space-time translations and rotations.  The $D$ generates dilatations, and the
$K_\mu$ generate the special conformal transformations (SCTs). These satisfy the commutation relations of the conformal algebra $so(4,2)$ of Minkowski space (with metric $\eta_{\mu \nu}={\rm diag}(-1,1,1,1)$),
\be
\[ \d_{P_\mu} , \d_{J_{\a \b}} \] &= \eta_{\mu \b} \d_{P_{\a}} - \eta_{\mu \a} \d_{P_{\b}},&
\[ \d_{J_{\a \b}} , \d_{J_{\g \d}} \] &= \eta_{\a \g} \d_{J_{\b \d}} - \eta_{\a \d} \d_{J_{\b \g}} + \eta_{\b \d} \d_{J_{\a \g}} - \eta_{\b \g} \d_{J_{\a \d}}, \nn \\
\[ \d_{P_\mu} , \d_{D} \] &= \d_{P_\mu},&
\[ \d_{J_{\a \b}} , \d_{K_{\mu}} \] &= \eta_{\a \mu} \d_{K_{\b}} - \eta_{\b \mu} \d_{K_{\a}}, \nn \\
\[ \d_{P_\mu} , \d_{K_\nu} \] &= 2 \eta_{\mu \nu} \d_D + 2 \d_{J_{\mu \nu}},&
\[ \d_D , \d_{K_{\mu}} \] &= \d_{K_{\mu}} \, ,
\label{conformalalgebra}
\ee
with all other commutators being zero.  

The Lagrangian must be such that the fields $\phi_I$ acquire a time-dependent background value
\be
\bar \phi_I (t) = {\alpha_I\over (-t)^{\Delta_I}}\,, \ \ \ \ \ \ \ \ \ \ \ \   -\infty<t<0\,,
\label{phibackgenmany}
\ee
where the $\alpha_I$'s are constant coefficients. To generate the desired symmetry breaking pattern, at least one field with 
weight $\Delta_I\not=0$ must have non-vanishing $\alpha_I$.  In this case, 10 of the 15 conformal generators~\eqref{delphiconfmany}
annihilate the background: $\delta_{P_i},\ \delta_D, \ \delta_{J_{ij}}, \ \delta_{K_i}\,, \ i=1,2,3.$
These 10 generators span an $so(4,1)$ sub-algebra, so this background realizes the symmetry breaking pattern \eqref{symmbreak}.

Expanding in fluctuations around the background~(\ref{phibackgenmany}), $\varphi _I=\phi_I -\bar\phi_I$, the unbroken generators act linearly on the perturbations $\varphi_I$,
\bea
\nonumber
\delta_{P_i}\varphi_I &=&-\partial_i\varphi_I \,, \qquad\,\,\,\,\,\,\,\,\,\,\,\,\,\,\,\,\,\,\,\,\,\,\,\,\,\,\, \delta_{J^{ij}}\varphi_I  =\(x^i\partial^j-x^j\partial^i\)\varphi_I \,, \\
\delta_D\varphi_I &=&-\(\Delta_I+ x^\mu\partial_\mu\)\varphi_I \,,\qquad \delta_{K_i}\varphi_I =\(-2x_i \Delta_I -2x_i x^\nu\partial_\nu +x^2\partial_i\)\varphi_I \,. \label{unbrokengenmanylin}
\eea
The broken generators act non-linearly, with a leading constant term (when the conformal weight is non-zero and the background is non-vanishing) plus a linear term,
\bea
\nonumber
\delta_{P_0}\varphi_I & =& {\Delta_I\over t}\bar\phi_I-\dot\varphi_I \,,\qquad \delta_{J^{0i}}\varphi_I =-{\Delta_I x^i\over t}\bar\phi_I+\(t\partial_i+x^i\partial_t\)\varphi_I \,, \\
\delta_{K_0}\varphi_I &=&-{\Delta_I x^2 \over t}\bar\phi_I+\(2t \Delta_I+ 2t x^\nu\partial_\nu +x^2\partial_t\)\varphi_I \,.
\label{brokengenmanylin}
\eea

As shown in~\cite{Hinterbichler:2011qk}, the symmetries alone determine much of the form of the quadratic action for the fluctuations, independent of the original Lagrangian. Invariance under the unbroken $so(4,1)$ subalgebra imposes that fields of different weights do not mix at the quadratic level, and that perturbations at high energies ({\it i.e.}, when their mass term can be ignored) propagate exactly at the speed of light. In the DBI case, however, we will instead find that perturbations travel
strictly subluminally. Imposing the non-linearly realized symmetries, the quadratic action for a given $\Delta$ is restricted to take the form
\be
S_{\Delta,{\rm quad}} \sim -\frac{1}{2}\int {\rm d}^4 x\bigg((-t)^{2(\Delta-1)}  \eta^{\mu\nu}\partial_\mu \varphi_I\partial_\nu \varphi_I  + M^{IJ} (-t)^{2(\Delta-2)} \varphi_I\varphi_J\bigg) \,.
\label{quadgen2}
\ee
If $\Delta \neq 0$, the mass matrix must satisfy the eigenvalue equation
\be
{M}^{IJ}\alpha_J= (\Delta + 1)(\Delta-4)\  \alpha^I \,,
\label{M3M1p}
\ee
whereas if $\Delta = 0$ it is unconstrained.

For example, if the theory has precisely one field $\phi$ of weight $\Delta\neq 0$, and this field has a non-vanishing background profile $\bar \phi(t) \sim 1/(-t)^\Delta$, then~(\ref{M3M1p})
determines its mass coefficient to be $M=(\Delta + 1)(\Delta-4)$. The quadratic action for the fluctuations of this field is thus fully determined (up to the overall normalization) by the symmetries to be
\be
S_{\rm quad}^{\Delta\neq 0}  \sim - \frac{1}{2}  \int {\rm d}^4 x\bigg((-t)^{2(\Delta-1)} \eta^{\mu\nu}\partial_\mu \varphi\partial_\nu \varphi +  (-t)^{2(\Delta-2)}(\Delta + 1)(\Delta-4)\varphi^2\bigg)\,.
\label{quadfinalsingle}
\ee
From this quadratic action we can determine the power spectrum. Transforming to the canonically normalized variable $v \equiv (-t)^{\Delta-1}\varphi$, the mode function equation becomes universal, independent of the
conformal weight $\Delta \neq 0$, 
\be
\ddot{v}_k  + \left(k^2 - \frac{6}{t^2}\right)v_k = 0,\qquad~~~~~~~ (\Delta\neq 0)\,. 
\label{veomlin}
\ee
The solution --- assuming the standard adiabatic vacuum initial condition --- is given by a Hankel function, $v_k  \sim {\sqrt{-t}}H_{5/2}^{(1)} (-kt).$
Outside the horizon, $k|t|\ll 1$, this gives the power spectrum $|v_k|^2 \sim \frac{1}{k^5t^4}$, or, in terms of the original field,
\be
P_\varphi(k) \sim \frac{1}{k^5t^{2(\Delta+1)}}\,,
\label{Pkred}
\ee
which is a strongly red-tilted spectrum. The scale invariant power spectrum required by observations does not come from this field but instead comes from an entropy field of approximate weight zero, as reviewed below. Nevertheless, a strongly red-tilted component is at first sight worrisome, since it would seem to dominate
any scale invariant contribution on sufficiently large scales. There are many ways to see that this component is in fact harmless.
In co-moving gauge, where $\varphi = 0$, the curvature perturbation acquires a strongly blue spectrum from the adiabatic mode,
$\zeta_k\sim 1/k^{3/2}$, which is negligible on large scales~\cite{Hinterbichler:2011qk}. In Newtonian gauge, the red-tilted
spectrum~(\ref{Pkred}) is an adiabatic perturbation which becomes a decaying mode in the standard, expanding
FRW phase~\cite{Creminelli:2010ba}.

The quadratic action also allows us to conclude that the background solution $\bar\phi \sim 1/(-t)^\Delta$ is a dynamical attractor.
Classically, the growing mode solution is
\be
\varphi \rightarrow \frac{1}{(-t)^{\Delta +1}}\,,
\ee
which can be re-summed into a harmless constant time-shift of the background solution:
\be
\bar{\phi}(t + \varepsilon)  = \bar{\phi}(t) +\varepsilon \dot{\bar{\phi}} (t)  \sim  \frac{1}{(-t)^{\Delta}} \left( 1 - \frac{\Delta \varepsilon}{t}\right)\,.
\label{phiattractgen}
\ee
Hence the perturbed field $\phi = \bar{\phi} + \varphi$ tends to the background solution, up to an irrelevant constant shift in time.\footnote{Note that this argument breaks down for many rolling fields --- an overall time shift may be used to remove the growing mode of a single field, but the others will generically diverge from the background solution at late times. In the special case where there is an $so(N)$ symmetry amongst the rolling fields, we may perform a rotation in field space so that there is a single adiabatic direction whose growing mode may be absorbed.}

The scale invariant spectrum which seeds structure formation in the late universe originates from perturbations of zero weight.
For weight-0 perturbations, the quadratic action~(\ref{quadgen2}) reduces to
\be
S_{\rm quad}^{\Delta = 0}  \sim- \frac{1}{2}  \int {\rm d}^4x \bigg(t^{-2}  \eta^{\mu\nu}\partial_\mu \vartheta_I\partial_\nu \vartheta^I  + M^{IJ} t^{-4} \vartheta_I\vartheta_J\bigg)\,.
\label{quadgenconf0v2}
\ee
The mass matrix $M^{IJ}$ is unconstrained, so the weight-0 fields generically have mass mixing. 
In the case of a single weight 0 field with a shift symmetry, this reduces to
\be
S_{\rm quad}^{(\Delta = 0)} \sim  - \frac{1}{2}  \int {\rm d}^4x\, t^{-2}  \eta^{\mu\nu}\partial_\mu \vartheta \partial_\nu \vartheta  \,,
\label{quadgenconf0v3}
\ee
which is exactly the action of a massless scalar on de Sitter space. Re-defining $u \equiv (-t) \vartheta$ the mode function equation is given by
\be
\ddot{u}_k +\left( k^2  - \frac{2}{t^2}\right)u_k = 0\,,
\label{fmodefcn}
\ee
with solution $u_k \sim \frac{e^{-ikt}}{\sqrt{2k}}\left(1-\frac{i}{kt}\right)$. The late time spectrum for $\vartheta$ is scale invariant,
\be
P_\vartheta(k) \sim {1\over 2k^3}\,.
\label{Pchi1}
\ee
This scale invariant entropy spectrum can subsequently be transferred to the adiabatic mode through well-known conversion
mechanisms~\cite{Lyth:2001nq,Dvali:2003em,Kofman:2003nx}. See~\cite{Wang:2012bq} for a discussion of conversion in the pseudo-conformal
and Galilean Genesis scenarios.

One of the simplest actions that can realize pseudo-conformal symmetry breaking is that of a massless scalar
field with a quartic potential~\cite{Craps:2007ch,Rubakov:2009np},
\be
S_\phi = \int \dfx \( -\half (\del \phi)^2 + \frac{\l}{4} \phi^4 \) .
\label{masslessquartic}
\ee
Classically, this is a conformal theory where $\phi$ has $\Delta=1$. For a spatially homogeneous field profile, the equation of motion is 
\be
\ddot{\phi} = \l \phi^3 \, .
\ee
Looking for a solution of the desired form $\phi = \alpha/(-t)^{}$, one finds that it only exists when $\lambda>0$, which in our convention corresponds to 
an upside-down quartic potential. The solution for $\alpha$ is then
\be
\alpha = \sqrt{\frac{2}{\l}}.
\label{linearbackground}
\ee
This is a zero energy solution, where the field starts from rest at the top of the potential in the asymptotic past, then rolls down. 
This solution is also a dynamical attractor~\cite{Craps:2007ch,Rubakov:2009np,Creminelli:2010ba,Hinterbichler:2011qk}. Coupling this sector minimally to gravity, one finds that the equation of state for
$\phi$ is always larger than unity during the phase of interest, $w \gg 1$, corresponding to a slowly contracting universe. This makes the
universe increasingly flat, homogeneous and isotropic~\cite{Gratton:2003pe,Hinterbichler:2011qk}, and hence addresses the
standard horizon and flatness problems which inflation was designed to solve. To generate scale-invariant perturbations, we
must couple in a spectator field $\vartheta$ of conformal weight-$0$ to the rolling field $\phi$. This can be achieved, for instance,
by promoting $\phi$ to a complex scalar field, with its radial part acquiring the $1/t$ background and its angular playing the
role of weight-0 perturbations~\cite{Rubakov:2009np}. Another incarnation of the scenario is Galilean Genesis~\cite{Creminelli:2010ba}, where the scalar field
action is that of the conformal galileon, which also admits a $1/t$ solution. 

We now turn the the main subject of this paper, extending the pseudo-conformal scenario to DBI-like non-linear realizations of the conformal algebra.

\section{DBI Pseudo-Conformal Scenario}
\label{dbisymmetrybreaking}
A `relativistic' extension of the pseudo-conformal mechanism can be obtained by considering the conformal DBI action \eqref{diracborninfeld}.
This action arises from the dynamics of a brane probing a bulk space-time.  The details of the construction can be found in~\cite{Goon:2011qf}, which we summarize here.
Consider an ambient higher-dimensional space-time with coordinates $X^A$ and metric $G_{A B}(X)$.
We consider a dynamical $3$-brane, with worldvolume coordinates $x^\mu$, probing this geometry.   The dynamical variables are the brane embedding functions $X^A(x)$, from which we construct the induced metric $g_{\mu\nu}^{\rm induced}(x)$ by pulling back the bulk metric
\bea \label{inducedmetricdef}
g_{\mu\nu}^{\rm induced}(x)&=& \frac{\partial X^A}{\partial x^\mu}\frac{\partial X^B}{\partial x^\nu}G_{AB}(X(x)) \ .
\eea
The induced metric transforms as a tensor under reparametrizations of the brane $\delta_g X^A=\xi^\mu\partial_\mu X^A$.  In addition to the gauge symmetry of reparametrization invariance, there can be global symmetries. Specifically, for every Killing vector $K^A(X)$ of the bulk metric there is a global symmetry under which the embedding scalars $X^A$ shift, $\delta_K X^A=K^A(X)$.  The induced metric~\eqref{inducedmetricdef} is invariant under these global symmetries. 

We choose to completely fix the reparameterization freedom by fixing the unitary gauge
\be
\label{ungaugechoice}
X^\mu(x)=x^\mu,\ \ \ \ \ \ \ \  X^I(x)\equiv\pi^I(x)\,,
\ee
where the subscript $I$ labels the co-dimensions of the brane. The world-volume coordinates of the brane are identified with the first four of the bulk coordinates.  The remaining unfixed fields, $\pi^I(x)$,
represent the transverse position of the brane in the higher-dimensional space. 

The form of the global symmetries is different once the gauge is fixed, because the gauge choice~(\ref{ungaugechoice}) is not in general preserved by these global symmetries.
The change induced by $K^A$ is $\delta_ Kx^\mu=K^\mu(x,\pi)$, $\delta_K\pi^I=K^I(x,\pi)$. To maintain the gauge~(\ref{ungaugechoice}), we must simultaneously perform a compensating gauge transformation with the gauge parameter $\xi_{\rm comp}^{\mu}=-K^\mu(x,\pi)$. The combined symmetry acting on the fields $\pi^I$ is now
\be
\label{gaugefixsym} 
(\delta_K+\delta_{{\rm comp}})\pi^I=-K^\mu(x,\pi)\partial_\mu\pi^I+K^I(x,\pi) \ ,
\ee
and will be a global symmetry of the gauge fixed action.  

We are interested in the case where the bulk space-time is AdS$_5$ with radius ${\cal R}$, in the Poincar\'e patch ,
\be
{\rm d}s^2_{\rm AdS} = G_{A B}{\rm d}X^A{\rm d}X^B={\cal R}^2\left[{1\over z^2}{\rm d}z^2+z^2\eta_{\mu\nu}{\rm d}x^\mu {\rm d}x^\nu\right] \,,
\ee 
where $0< z< \infty$ is the radial AdS direction. In addition to the manifest Poincar\'e symmetries of the $x^\mu$ coordinates,
AdS$_5$ has five additional Killing vectors
\bea 
K_\mu&=&2x_\mu z\partial_z+\left({1\over z^2}+x^2\right)\partial_\mu-2x_\mu x^\nu\partial_\nu\,, \nn\\
D&=&-z\partial_z+x^\mu\partial_\mu\,. 
\label{adskilling}
\eea
There will be one transverse $\pi^I$ field corresponding to the radial direction $z$, and we will call this field $\phi$.  According to~\eqref{gaugefixsym}, these generate the following global symmetries on $\phi$ in the gauge-fixed action,
\bea  
\delta_{D}\phi&=&-\( \D_\phi + x^\nu \del_\nu \) \phi\,,\nn \\
\delta_{K_\mu}\phi&=& - 2 x_\mu \( \D_\phi + x^\nu \del_\nu \) \phi + x^2 \del_\mu \phi + \frac{1}{\phi^2} \del_\mu \phi\,, 
\label{adssym1}
\eea
where $\D_\phi = 1$. In addition, the manifest Poincar\'e symmetries of the $x^\mu$ coordinates generate the standard Poincar\'e transformations on $\phi$.
Together, the 5 symmetries~\eqref{adssym1} and the 10 Poincar\'e symmetries satisfy the algebra~\eqref{conformalalgebra} and provide a non-linear realization of $so(4,2)$.
Compared to the transformations~\eqref{delphiconfmany} in the standard case, there is an extra term $\phi^{-2} \del_\mu \phi$ in the expression for $\d_{K_\mu} \phi$;
in the DBI action, the special conformal transformations are thus realized non-linearly.

To construct the leading order action for the brane, we combine a tadpole potential term with a kinetic term arising from the induced volume form on the brane.  
The induced metric on the brane~(\ref{inducedmetricdef}) is, in the gauge \eqref{ungaugechoice},
\be
g_{\mu\nu}^{\rm induced}(x) = \R^2 \phi^2 \( \eta_{\mu \nu} + \frac{\del_\mu \phi}{\phi^2} \frac{\del_\nu \phi}{\phi^2} \)
\, ,
\label{vol}
\ee
hence the world-volume action arising from the determinant is
\be
S_2\equiv \R^{-4}  \int \dfx \sqrt{-g^{\rm induced}} =  \int \dfx  \frac{\phi^4}{\gamma} \, ,
\label{volaction}
\ee
where we have introduced the Lorentz factor
\be
\g \equiv \frac{1}{\sqrt{1+(\del \phi)^2/\phi^4}}\, ,
\label{gamdef}
\ee
and where indices are contracted with $\eta_{\mu\nu}$. 
Meanwhile, the tadpole action
\be
S_1 \equiv \int \dfx \phi^4
\label{tad}
\ee
is the unique local action which does not depend on derivatives and is invariant under all 15 of our symmetries.  Geometrically, it is the proper 5-volume between the brane and some fixed reference brane~\cite{Goon:2011qf}.

Combining the tadpole~(\ref{tad}) and induced volume form~(\ref{vol}), with a relative coefficient governed by $\lambda$, we arrive at the DBI action.
\be
S_{\rm DBI} = \left(1 + \frac{\lambda}{4}\right) S_1 - S_2 = \int \dfx \phi^4 \( 1 + \frac{\l}{4} -\gamma^{-1}  \) .
\label{dbiaction}
\ee
For convenience we have chosen the constant so that a Poincar\'e invariant solution $\phi={\rm constant}$ exists only when $\lambda=0$, and have normalized the action so that expanding around this solution we have a canonical, healthy scalar kinetic term.  Note that in the limit of small field gradients, $|(\del\phi)^2|\ll \phi^4$, this action reduces to the negative quartic model~(\ref{masslessquartic}).

A field configuration $\phi=f$, where $f$ is a constant, is preserved only by the Poincar\'e subalgebra spanned by $\d_{P_\mu},\ \d_{ J_{\mu \nu} }$.  Expanding in fluctuations about such a configuration, $\phi=f+\varphi$, the action of the symmetry generators on $\varphi$ is 
\bea
\d_{P_\mu} \varphi &=& -\del_\mu \varphi \,, \qquad  \;\;\d_{K_\mu} \varphi = -2x_\mu \D_\phi f- 2 x_\mu \( \D_\phi + x^\s \del_\s \) \varphi + x^2 \del_\mu \varphi + \frac{1}{(f+\varphi)^2} \del_\mu \varphi, \, 
\nn \\
\qquad\d_{ J_{\mu \nu} } \varphi &=& x_\mu \del_\nu \varphi - x_\nu \del_\mu \varphi \,,~~~~~~~\d_D \varphi = -\D_\phi f- \( \D_\phi + x^\mu \del_\mu \) \varphi\,,
\label{nonlineargeneratorsphi}
\eea
A symmetry is broken if and only if the transformation acting on the fluctuation has a constant part.  The only transformations without a constant part are the Poincar\'e transformations $\d_{P_\mu},\ \d_{ J_{\mu \nu} }$, so we confirm that the symmetry breaking pattern is $so(4,2)\rightarrow$ Poincar\'e in this case. The difference here is that the special conformal transformations are now non-linear, since there are quadratic and higher order pieces coming from expanding out $\frac{1}{(f+\varphi)^2}$ in powers of the fluctuation.  In Sec.~\ref{linearpseudoconformal}, the transformations were all at most linear in the fluctuations.

Looking for purely time-dependent solutions, $\phi = \pb(t)$, the equation of motion derived from~(\ref{dbiaction}) reduces to
\be
\ddt \( \bar{\g} \, \dot{\pb} \) = \pb^3 \( 4 + \l - 2 \bar{\g}^{-1} - 2 \bar{\g} \) \, ,
\label{dbieom}
\ee
where $\bar{\g} = 1/\sqrt{1 - \dot{\pb}^2/\pb^4} \geq 1$. We look for solutions of the form
\be
\pb(t) = \frac{\a}{(-t)}
\nc
-\infty < t < 0\,,
\label{dbibackground}
\ee
where $\alpha$ can be assumed positive without loss of generality since the theory is $Z_2$ symmetric.
On the background~(\ref{dbibackground}), the relativistic factor $\g$ is a constant, 
\be
\bar{\g}(\alpha) = \frac{1}{\sqrt{1 - 1/\a^2}} >1 \, ,
\label{gamalpha}
\ee
and the equation of motion~(\ref{dbieom}) becomes
\be 
\bar{\gamma}(\alpha) = 1 + \frac{\l}{4} \,.
\label{gamalphaeom}
\ee
In the ``non-relativistic" limit, $\alpha \gg 1$, we recover the relation~(\ref{linearbackground}) between $\alpha$ and $\lambda$. 
More generally, since $\gamma\geq 1$ the existence of a non-trivial solution requires 
\be 
\lambda>0\,.
\ee

The solution~\eqref{dbibackground} is annihilated by the 10 generators $\d_D$, $\d_{P_i}$, $\d_{K_i}$, and $\d_{J_{i j}}$, but not by the 5 generators $\d_{P_0}$, $\d_{K_0}$, or $\d_{J_{0 i}}$,
which act as
\be
\d_{P_0} \pb = \frac{\pb}{t} \, ; \qquad \d_{ J_{0 i} } \pb  = \frac{x^i \pb}{t} \,; \qquad \d_{K_0} \pb = - \( x^2 + \frac{1}{\pb^2} \) \frac{\pb}{t} \, .
\label{brokenact}
\ee
Our background therefore spontaneously breaks the $so(4,2)$ symmetry of the DBI action down to an $so(4,1)$ subalgebra, realizing pseudo-conformal symmetry breaking in the same manner as the background~\eqref{linearbackground}.  

\subsection{Quadratic action for fluctuations}

For perturbations $\f \equiv \phi - \pb$ about the background scaling solution~\eqref{dbibackground}, the unbroken $so(4,1)$ subalgebra action starts at linear order in $\f$,
\be
\d_{P_i} \f &= -\del_i \f & \d_{J_{i j}} \f &= \( x^i \del_j - x^j \del_i \) \f \nn \\
\d_{D} \f &= - \(1 + x^{\mu} \del_{\mu} \) \f& \d_{K_i} \f &= - 2 x^i \f - 2 x^i x^{\l} \del_{\l} \f + \( x^2+ \frac{1}{\pb^2} \) \del_i \f +{\cal O}(\varphi^2)\, ,
\label{DBIquad0}
\ee
while the broken generators start at zeroth order in $\f$,
\be
\d_{P_0} \f = \frac{\pb}{t}+{\cal O}(\varphi)
\nc
\d_{J_{0 i}} \f = x_i \frac{\pb}{t}+{\cal O}(\varphi)
\nc
\d_{K_0} \f = - \( x^2 + \frac{1}{\pb^2} \) \frac{\pb}{t}+{\cal O}(\varphi)
\, .
\label{DBIquad0b}
\ee
The difference with the transformations here and those of~\eqref{unbrokengenmanylin} and~\eqref{brokengenmanylin} in Sec.~\ref{linearpseudoconformal} is that the transformations now contain higher order pieces from expanding the denominators, even though the symmetry breaking pattern is the same. In addition, the transformations at linear order for the unbroken generators, and at zeroth order for the broken generators, are different because of the $1/\bar\phi^2$ terms. 

These differences in the transformation rules result in perturbations having a strictly subluminal sound speed. 
Expanding~\eqref{dbiaction} around the background solution~\eqref{dbibackground}, the quadratic action for the fluctuations $\f$ is
\be
S = \half \bar{\g}^3 \int \dfx \( \dot{\f}^2 - \frac{1}{\bar{\g}^2}(\del_{i} \f)^2 + \frac{6}{t^2} \f^2 \) \,. 
\label{dbiquad}
\ee
As advocated, the sound speed of the fluctuations is strictly less then one
\be
c_s = \frac{1}{\bar{\g}} < 1 \, .
\ee
More generally, we will see in Sec.~\ref{genquadDBI} that the quadratic action~(\ref{dbiquad}), and in particular the sound speed, is completely fixed by the symmetries (except for the overall normalization).

Using this quadratic action, we can compute the power spectrum of $\f$. In terms of the sound horizon time $y \equiv t/\bar{\g}$ and the canonically
normalized variable $v \equiv \bar{\g} \f$, the mode function equation takes the same form as in the luminal case~(\ref{veomlin}):
\be
v''_k + \( k^2 - \frac{6}{y^2} \) v_k = 0\,,
\ee
where $' \equiv \del/\del y$. As before, the power spectrum is strongly tilted to the red:
\be 
P_\varphi(k)={9\over 2}{\bar{\gamma}^2 \over k^5(-t)^4}\,.
\label{PvarphiDBI}
\ee
The scale invariant contribution must once again arise from weight-0 entropy perturbations. In Sec.~\ref{dbiscaleinvariance} we will show how
these can naturally arise as embedding coordinates for a brane moving in additional co-dimensions. 

By examining the behavior of the perturbations, we can see that the background~\eqref{dbibackground} is again a dynamical attractor.
In the limit $k\rightarrow 0$, where spatial gradients can be neglected, the quadratic action~(\ref{dbiquad}) agrees with~(\ref{quadfinalsingle}) 
with $\Delta = 1$. Thus the time-dependence of the growing mode is identical, and the time-shift argument~(\ref{phiattractgen}) carries over to the DBI case.


\section{Another Example: DBI Galilean Genesis}
\label{conformalDBIgalileon}

The original Galilean Genesis scenario \cite{Creminelli:2010ba} relies on the conformal galileon terms, which consist of conformally-invariant derivative interaction terms,
to generate a $1/t$ background. The stress energy tensor of the conformal galileon can violate the null energy condition without ghost instabilities
or other pathologies~\cite{Nicolis:2009qm}, and thus can drive an expanding phase from an asymptotically static past --- the Galilean Genesis solution. One drawback of the scenario, however, is that although perturbations propagate exactly at the speed of light on this background --- as dictated by the general symmetry analysis reviewed in Sec.~\ref{linearpseudoconformal} ---
perturbations can propagate superluminally on slightly different backgrounds, which lie within the purview of the effective theory. 

In this Section we consider the DBI generalization of the Galilean Genesis scenario~\cite{Creminelli:2010ba}. Aside from offering another example of
our symmetry-breaking pattern, it also presents a cure to the superluminality problem --- perturbations must propagate subluminally on the
$1/t$ background --- while retaining the same number of symmetries. Alternatively, the sound speed can be made subluminal through
explicit breaking of special conformal transformations~\cite{Creminelli:2012my}.

The action of the conformal DBI galileon has five independent conformally-invariant terms \cite{deRham:2010eu, Goon:2011qf}. For simplicity, we focus on the first three terms: 
\be
S = \int{\rm d}^4x\Big(c_1 {\cal L}_1 + c_2 {\cal L}_2 + c_3 {\cal L}_3\Big)
\, ,
\label{galileonaction}
\ee
where the three conformal DBI galileon Lagrangians are given by
\be
&
{\cal L}_1 =  \phi^4\,;
\nl
{\cal L}_2 =  \frac{ \phi^4}{\g}\,;
\nl
{\cal L}_3 =  \g^2 \frac{\del_\mu \phi \, \del_\nu \phi\, \del^\mu \del^\nu \phi}{\phi^3} + \phi^4 \( \frac{1}{\g^2} + 5 - 2 \g^2 \)  \,.
\ee
The ${\cal L}_1$ and ${\cal L}_2$ contributions have been discussed already and correspond respectively to a tadpole~(\ref{tad}) and the invariant volume of the induced metric~(\ref{vol}).
The term ${\cal L}_3$ comes from considering the extrinsic curvature on the brane~\cite{deRham:2010eu,Goon:2011qf}. As a special case of the above action,
the DBI action discussed earlier in~\eqref{diracborninfeld} is recovered by setting 
\be
c_1 = 1 + \frac{\lambda}{4}\,;~~ c_2 = -1\,;~~c_3 = 0\qquad ({\rm DBI}~{\rm Pseudo-Conformal})\,.
\label{DBIpseudo}
\ee
Another case of interest is the DBI extension of the Galilean Genesis scenario. To parallel the original Galilean Genesis~\cite{Creminelli:2010ba},
which only relied on derivative interactions for the scalar, we impose that the tadpole around a $\phi={\rm constant}$ configuration vanishes. This requires:
\be
c_1 + c_2+4c_3=0 \qquad ({\rm DBI}~{\rm Galilean}~{\rm Genesis})\,.
\label{DBIGen}
\ee

For a time-dependent background, $\phi = \pb(t)$, the equation of motion is
\be
4 c_1
-c_2 \bar{\g}^3 \( - \frac{\ddot{\pb}}{\pb^3} + 6 \frac{\dot{\pb}^2}{\pb^4} - 4 \)
+c_3 \bar{\g}^4 \( 6 \frac{\ddot{\pb}}{\pb^3} + 4 \frac{\dot{\pb}^4}{\pb^8}
- 32 \frac{\dot{\pb}^2}{\pb^4} + 16  \)  =0 \,.
\ee
Focusing on the desired scaling ansatz, $\phi = \alpha/(-t)$, this reduces to
\be
0 = c_1 + c_2\bar{\g}(\alpha) + c_3 \( 1 + 3 \bar{\g}^2(\alpha) \) \,,
\label{galileongammaequation}
\ee
where $\bar{\g}(\alpha)$ is again given by~(\ref{gamalpha}). This fixes $\alpha$ in terms of the coefficients $c_1$, $c_2$ and $c_3$.
As it should, this equation reproduces~(\ref{gamalphaeom}) for the DBI pseudo-conformal parameters~(\ref{DBIpseudo}). 
More generally, if all three coefficients are non-zero then there are potentially two solutions:
\be 
\bar{\g}_{\pm}(\alpha) = \frac{-c_2 \pm \sqrt{ c_2^2 - 12 c_3 \( c_1 + c_3 \) } }{6 c_3}\,.
\label{gammapm}
\ee
Clearly, for either potential solution to be real we must require $c_2^2 \geq 12 c_3 \( c_1 + c_3 \)$. Moreover, we need either $\g_->1$ or $\g_+>1$ to have a physically allowed solution.

In the limit $c_3\rightarrow 0$, only one of these solutions matches continuously onto the pure DBI solution found in Sec.~\ref{dbisymmetrybreaking}, while the other is a new branch which is not analytic in $c_3$. 
Specifically, if $c_2< 0$, then we have $\gamma_+\simeq -c_2/3 c_3$, $\gamma_- \simeq -c_1/c_2$ for small $c_3$, hence it is the ``$-$" branch which matches smoothly with the pure DBI theory in this case; if $c_2>0$,
we instead have $\gamma_+\simeq -c_1/c_2$, $\gamma_-\simeq  -c_2/3 c_3$, and now it is the ``$+$" branch which matches continuously to the DBI solution.

\subsection{Quadratic action for fluctuations}

Next we consider the stability of this solution. Expanding $\phi$ about the background as $\phi = \pb + \f$, the quadratic action for the fluctuations is
\be
S_{\rm quad} = \mp \frac{1}{2} \bar{\g}_\pm^3(\alpha) \sqrt{  c_2^2 - 12 c_3 \( c_1 + c_3 \)  } \int \dfx \( \dot{\f}^2 - \frac{1}{\bar{\g}_\pm^2(\alpha)} (\del_i \f)^2 + 6 \frac{\phi^2}{t^2} \) \,.
\ee
The square root factor out front is always positive, due to the requirement that a solution exists, {\it i.e.} that $\bar{\gamma}_\pm$ is real. Moreover, since $\bar{\gamma}_\pm > 1$ for physically allowed solutions,
we see from the overall $\mp$ sign multiplying the quadratic action that fluctuations around the $\bar{\gamma}_-$ branch are necessarily stable, while perturbations around the $\bar{\gamma}_+$ branch
are necessarily unstable. In what follows it will helpful to also consider the stability round a trivial $\bar{\phi} = {\rm constant}$ solution. The kinetic term around such a background is
\be
{\cal L}_{\rm kinetic} \sim {1\over 2}\left(6c_3 + c_2\right) \int \dfx (\partial\varphi)^2\,.
\ee

Let us first specialize to a small deformation of the pure DBI case studied in Sec.~\ref{dbisymmetrybreaking}. Namely, we set $c_1 = T(1 + \lambda/4)$, $c_2 = -T$, where we have allowed for an overall coefficient $T$,
and consider the limit of small $c_3$. When $T > 0$, corresponding to a positive-tension brane, the trivial background $\bar{\phi} = {\rm constant}$ is stable, while fluctuations around the time-dependent solution are only stable around the $\bar{\gamma}_-$ branch which matches smoothly to the DBI theory; fluctuations are unstable around the non-analytic $\bar{\gamma}_+$ branch. When $ T < 0$, on the other hand, the trivial background is unstable,
and so is the $\bar{\gamma}_+$ branch which matches smoothly to the pure DBI theory; instead it is new branch that is stable in this case.

Let us now focus on the DBI Galilean Genesis case~(\ref{DBIGen}), corresponding to $c_1 + c_2+4c_3=0$. There is at most only one scaling solution in this case, with
\be
\bar{\gamma}_{\rm Genesis} = -\frac{c_2}{3c_3}-1\,.
\label{gamgen}
\ee
If $6c_3 + c_2 < 0$, for which the $\bar{\phi} = {\rm constant}$ solution is stable, the DBI Genesis solution only exists ({\it i.e.}, $\gamma > 1$) if $c_2<0$, and it corresponds to the $\bar{\gamma}_+$ branch
--- fluctuations around the time-dependent background are therefore unstable in this case. If $6c_3 + c_2 > 0$, for which the trivial solution is unstable, the DBI Genesis solution only exists for $c_2 > 0$,
and it corresponds to the $\bar{\gamma}_-$ branch --- fluctuations around the time-dependent background are stable in this case. Thus the Genesis and trivial backgrounds always have opposite
stability, just as in the original Genesis scenario~\cite{Creminelli:2010ba}. (This need not be the case if we include all 5 DBI conformal galileon terms. We leave a study of this general case to future work~\cite{DBIfuture}.)


\section{Weight Zero Fields and Scale Invariance: An Example from  AdS$_5 \times$S$^1$ Brane Embedding}
\label{dbiscaleinvariance}

The DBI pseudo-conformal framework has the advantageous property that weight-0 fields, necessary for generating scale invariant perturbations, have
a natural geometric interpretation in terms of the brane moving in additional co-dimensions. 
As in Sec. \ref{dbisymmetrybreaking}, we consider a 3-brane probing a higher-dimensional geometry. However, we now take the higher-dimensional space to be a product
${\cal M} = {\rm AdS}_5\times{\cal Y}$, where ${\cal Y}$ is a compact manifold of arbitrary dimension, $n$. We choose coordinates so that the first 5 coordinates $X^a$ parameterize the anti-de Sitter space and the other $n$ coordinates $y^I$ parameterize the internal manifold. The line element associated with the bulk metric then takes the form
\be
G_{AB}{\rm d}X^A{\rm d}X^B = g_{ab}^{\rm Ads}{\rm d}X^a{\rm d}X^b+h_{IJ}{\rm d}y^I{\rm d}y^J~,
\ee
where $h_{IJ}(y)$ is the metric on the compact space ${\cal Y}$. We fix diffeomorphism invariance by choosing the gauge
\be
X^\mu (x)= x^\mu~,~~~~~~~~~~~X^5(x) \equiv \phi(x)~,~~~~~~~~~~~y^I(x) \equiv \Phi^I(x)~.
\ee
As before, we construct diffeomorphism scalars from the induced metric and its curvature invariants. In addition to the global symmetries of the action inherited from the ${\rm AdS}_5$ 
the action will have global symmetries inherited from the Killing vectors of the internal manifold.  
In the case that these global symmetries appear like shift symmetries of the fields, we can expect the fields to acquire a scale-invariant spectrum of perturbations.

Let us illustrate the idea with the simplest case where the compact space is a circle.  Thus we consider a dynamical $3$-brane probing a bulk space-time ${\cal M}={\rm AdS}_5\times S^1$, consisting of AdS$_5$ in the Poincar\'e patch with radius ${\cal R}$, times a circle of radius $\ell$,
\be
{\rm d}s^2 = G_{A B}{\rm d}X^A{\rm d}X^B={\cal R}^2\left[{1\over z^2}{\rm d} z^2+z^2\eta_{\mu\nu}{\rm d}x^\mu {\rm d}x^\nu\right]+\ell^2 {\rm d}\Theta^2\,,
\ee
where the $A,B$ indices now run from $0$ to $5$, and $0 < \Theta < 2\pi$ is an angular coordinate for the $S^1$. Fixing unitary gauge, as we did in~(\ref{ungaugechoice}), there
are now two fields $\phi$ and $\theta$, which represent the transverse position of the brane in the radial AdS direction and in the $S^1$, respectively:
\be
\label{ungaugechoice2}
X^\mu(x)=x^\mu\,,~~~~~~~ X^5(x)\equiv\phi(x)\,,~~~~~~~X^6\equiv \theta (x) \ .
\ee

According to \eqref{gaugefixsym}, we have the following AdS global symmetries on $\phi$ and $\theta$ in the gauge fixed action
\bea  
\nonumber
\delta_{D}\phi&=&-\( \D_\phi + x^\nu \del_\nu \) \phi\,;\qquad \delta_{K_\mu}\phi =  - 2 x_\mu \( \D_\phi + x^\nu \del_\nu \) \phi + x^2 \del_\mu \phi + \frac{1}{\phi^2} \del_\mu \phi \,;\\
\delta_{D}\theta&=&-\( \D_\theta + x^\nu \del_\nu \) \theta\,; \qquad  \delta_{K_\mu}\theta = - 2 x_\mu \( \D_\theta + x^\nu \del_\nu \) \theta + x^2 \del_\mu \theta + \frac{1}{\phi^2} \del_\mu \theta\,,
\eea
where $\D_\phi = 1$ and $\D_\theta = 0$.  The manifest Poincar\'e symmetries of the $x^\mu$ coordinates then generate the standard Poincar\'e transformations on $\phi$ and $\theta$.
In addition to the Poincar\'e generators and the AdS$_5$ Killing vectors~\eqref{adskilling}, there is also the Killing vector generating translations along the $S^1$,
\be 
C=\partial_\Theta\,.
\ee
The action of the $S^1$ generator on $\phi$ is trivial, $\d_C \phi = 0$, while its action on $\theta$,
\be
\d_C \theta = 1 \, ,
\ee
corresponds to a shift symmetry.  This is exactly the extra symmetry we will need to protect the scale invariance of $\theta$ perturbations. The 15 AdS$_5$ generators satisfy the algebra~\eqref{conformalalgebra}, while the $S^1$ generator $\d_C$ commutes with itself and all of the AdS$_5$ generators.

To construct the leading order action for the brane, once again we combine a tadpole potential term with a kinetic term arising from the induced volume form on the brane.  
The induced metric on the brane is once again given by~(\ref{inducedmetricdef}). In the unitary gauge~(\ref{ungaugechoice2}), it takes the form
\be
g_{\mu \nu}^{\rm induced} (x) = \R^2 \phi^2 \( \eta_{\mu \nu} + \frac{\del_\mu \phi}{\phi^2} \frac{\del_\nu \phi}{\phi^2} +{\ell^2\over{\cal R}^2}\frac{\del_\mu \theta}{\phi} \frac{\del_\nu \theta}{\phi} \) \,,
\, 
\ee
hence the volume form arising from the determinant is given by
\be
\R^{-4} \sqrt{-\bar g} = \phi^4 \sqrt{1 + \frac{(\del \phi)^2}{\phi^4} + {\ell^2\over{\cal R}^2}\frac{(\del \theta)^2}{\phi^2}
+{\ell^2\over{\cal R}^2} \frac{(\del \phi)^2 (\del \theta)^2 - (\del \phi \cdot \del \theta)^2}{\phi^6}} \, .
\ee
Meanwhile, the tadpole action~(\ref{tad}) is the unique local action which does not depend on derivatives and is invariant under all 16 symmetries of the AdS$_5 \times S^1$ construction.
(In particular, invariance under the shift symmetry $\theta \gt \theta + c$ implies that the tadpole does not depend on $\vartheta$.) Thus we consider the following action:
\be
S_{\phi \theta} = \int \dfx \phi^4 \( 1 + \frac{\l}{4} - \sqrt{1 + \frac{(\del \phi)^2}{\phi^4} + \frac{(\del \theta)^2}{\phi^2}
+ \frac{(\del \phi)^2 (\del \theta)^2 - (\del \phi \cdot \del \theta)^2}{\phi^6}} \)\,,
\label{fullaction}
\ee
where we have canonically normalized $\theta$ so that it now ranges over $(0,{2\pi\ell\over{\cal R}})$.

\subsection{Pseudo-conformal background}

To realize pseudoconformal symmetry breaking with this action, we must show that the equations of motion admit a solution for which
\be
\pb = \frac{\alpha}{(-t)^{\D_\phi}} = \frac{\alpha}{(-t)} \,;
\qquad \bar{\theta} = \frac{\theta_0}{(-t)^{\D_\theta}} = \theta_0\,,
\label{fullscaling}
\ee
where $\alpha >0$, without loss of generality, and $\theta_0$ is constant (which can be arbitrary, thanks to the shift symmetry).
For purely time-dependent field profiles, the equations of motion are
\be
\frac{\rm d}{{\rm d}t} \( \bar{\g} \dot{\bar{\phi}} \)
=\bar{ \g} \phi \dot{\bar{\theta}}^2 + \bar\phi^3 \( 4 + \l - 2 \bar{\g} - \frac{2}{\bar{\g}} \)\,;\ \ \ \ \
\frac{\rm d}{{\rm d}t} \( \bar{\g} \bar{\phi}^2 \dot{\bar{\theta}} \) = 0 \, ,
\label{phichieom}
\ee
where the background relativistic factor is 
\be
\bar{\g} \equiv \frac{1}{\sqrt{1 - \dot{\pb}^2/\pb^4 - \dot{\bar{\theta}}^2/\pb^2}} \,.
\ee
Substituting our ansatz~(\ref{fullscaling}), the equation of motion implies that $\bar{\g}$ is constant and related to $\lambda$ by
\be
\bar{\g}(\alpha) = \frac{1}{\sqrt{1 - 1/\a^2}} = 1 + \frac{\l}{4} \, . \label{fullalpha}
\ee

The background $\pb = - \a/t$ is annihilated by the generators $\d_D$, $\d_{P_i}$, $\d_{K_i}$, and $\d_{J_{i j}}$, as well as $\d_C$, but not by the 5 generators $\d_{P_0}$, $\d_{K_0}$, or $\d_{J_{0 i}}$, which act as~(\ref{brokenact}). The background $\bar{\theta} = \theta_0$ is annihilated by all 15 conformal generators $\d_{P_\mu}$, $\d_{J_{\mu \nu}}$, $\d_D$, $\d_{K_\mu}$, but not by the shift symmetry $\d_C$, which acts as $\d_C \bar\theta = 1$.
The background solution $\pb=-\a/t$, $\bar{\theta}=\theta_0$ therefore spontaneously breaks six of the 16 symmetries of the action.  The 10 unbroken generators
$\d_{P_i}
\, , 
\d_{J_{i j}}
\, , 
\d_D
\, , 
\d_{K_i}
$
generate a residual $so(4,1)$ algebra.  The background~\eqref{fullscaling} realizes pseudo-conformal symmetry breaking~(\ref{symmbreak}), and also spontaneously breaks the shift symmetry $\theta \gt \theta + c$.

\subsection{Quadratic action for fluctuations}
\label{quadAdS5S1}

Consider now how the symmetries act on the fluctuations $\phi = \bar{\phi} + \varphi$ and $\theta = \bar{\theta} + \vartheta$.
To leading order in $\f$, the unbroken $so(4,1)$ subalgebra acts linearly on $\f$,
\be
\d_{P_i} \f &= -\del_i \f\,, & \d_{J_{i j}} \f &= \( x^i \del_j - x^j \del_i \) \f\,, \nn \\
\d_{D} \f &= - \( \D_\phi + x^{\mu} \del_{\mu} \) \f \,,& \d_{K_i} \f &=  - 2 x_i \( \D_\phi + x^{\mu} \del_{\mu} \) \f + \( x^2+ \frac{1}{\pb^2} \) \del_i \f + \ldots \, \,,
\label{DBIquad1}
\ee
while the broken conformal generators act non-linearly on $\f$,
\be
\d_{P_0} \f = \frac{\pb}{t} +\ldots 
\nc
\d_{J_{0 i}} \f = x_i \frac{\pb}{t} +\ldots 
\nc
\d_{K_0} \f = - \( x^2 + \frac{1}{\pb^2} \) \frac{\pb}{t} +\ldots 
\, .
\label{DBIquad2}
\ee
The ellipses in these expressions indicate terms that serve to constrain contributions to the action of higher than quadratic order, and hence can be ignored for the present purpose. 
To leading order in $\f$, the unbroken $so(4,1)$ subalgebra acts linearly on $\vartheta$,
\be
\d_{P_i} \vartheta &= -\del_i \vartheta\,, &
\d_{J_{i j}} \vartheta &= \( x^i \del_j - x^j \del_i \) \vartheta\,, \nn \\
\d_{D} \vartheta &= - \( \D_\theta + x^{\mu} \del_{\mu} \) \vartheta \,,&
\d_{K_i} \vartheta &= - 2 x^i \( \D_\theta + x^{\mu} \del_{\mu} \) \vartheta + \( x^2+ \frac{1}{\pb^2} \) \del_i \vartheta +\ldots 
\label{DBIquad3}
\ee
while the broken conformal generators also act linearly on $\vartheta$,
\be
\d_{P_0} \vartheta &= - \del_t \vartheta
\nc
\d_{J_{0 i}} \vartheta = t \del_i \vartheta - x_i \del_t \vartheta \,,
\nn \\
\d_{K_0} \vartheta &= - 2 t \( \D_\theta + x^{\mu} \del_{\mu} \) \vartheta + \( x^2+ \frac{1}{\pb^2} \) \del_t \vartheta + \ldots
\label{DBIquad4}
\ee
The shift symmetry acts on $\f$ and $\vartheta$ as
\be
\d_C \f = 0
\nc
\d_C \vartheta = 1
\, .
\ee

Expanding the action~(\ref{fullaction}) around the background $\bar{\phi} = \alpha/(-t)$ and $\bar{\theta} = \theta_0$ to quadratic order in the perturbations $\varphi$ and $\vartheta$,
we obtain
\be
S_{\rm quad}
&
= \half \bar{\g}^3 \int \dfx \( \dot{\f}^2 - \frac{1}{\bar{\g}^2}(\del_i \f)^2 + \frac{6}{t^2} \f^2 \)
+ \half \bar{\g} \int \dfx \pb^2(t) \( \dot{\vartheta}^2 - \frac{1}{\bar{\g}^2}(\del_i \vartheta)^2 \) \,.
\label{quadactiongeom}
\ee
Both $\f$ and $\vartheta$ propagate with identical, subluminal sound speed $c_s= \bar{\gamma}^{-1} <1$. 
Since these fields have different weights, they do not mix at quadratic level, consistent with the general discussion in Sec.~\ref{genquadDBI}.
The $\varphi$ part of the action is identical to~(\ref{dbiquad}), and hence leads to the same power spectrum as in~(\ref{PvarphiDBI}).
To calculate the power spectrum for $\vartheta$, we introduce as before the sound horizon time $y \equiv t/\bar{\g}$ and define the canonically normalized variable
$u \equiv \pb \vartheta$:
\be
S_\vartheta
&
= \half \int \dy \dcx \( (u')^2 - (\del_i u)^2 + \frac{2}{y^2} u^2 \) .
\label{normfullquad}
\ee
The equation of motion for the mode functions is given by
\be
u''_k + \( k^2 - \frac{2}{y^2} \) u_k = 0 
\,,
\ee
whose solution with the standard adiabatic vacuum is $u_k = \frac{e^{- i k y}}{\sqrt{2k}} \( 1 - \frac{i}{k y} \)$. 
The power spectrum for the original variable is therefore scale invariant:
\be 
P_\vartheta(k)={\bar{\gamma}^2-1\over 2}{1\over k^3} \,.
\ee
Note that the amplitudes of the fields $\vartheta$ and $\varphi$ are related: they are both set by $\bar\gamma$.


\section{The General Quadratic Action}
\label{genquadDBI}

In this Section, we apply symmetry arguments to derive the most general 2-derivative quadratic action for perturbations around the background~(\ref{phibackgenmany}),
including multiple fields $\phi_I$ of arbitrary conformal weights $\Delta_I$. The derivation closely parallels that presented in~\cite{Hinterbichler:2011qk} (see Sec.~2 of that paper) and reviewed in Sec.~\ref{linearpseudoconformal}, with the key difference being that the speed of propagation is now fixed to be subluminal, because the conformal symmetries of interest are of the DBI type, with special conformal transformations including terms of all orders in $\varphi _I=\phi_I -\bar\phi_I$.
We will see that the resulting action is fixed by the symmetries up to a few constants. The action will linearly realize the unbroken symmetries, and
non-linearly realize the broken ones.

For the purpose of deriving the quadratic action, the main differences between the DBI conformal symmetries --- {\it e.g.},~(\ref{DBIquad1})$-$(\ref{DBIquad4}) ---
and the ordinary ones --- {\it e.g.},~(\ref{brokengenmanylin}) --- amount to additional $1/\bar{\phi}^2 \sim t^2$ contributions to the $\delta_{K_\mu}$ transformations.
Inspired by the transformation rules~(\ref{DBIquad1})$-$(\ref{DBIquad4}) that arise in the geometric construction, we assume the unbroken symmetries act on the
fluctuations linearly as 
\be
\d_{P_i} \f_I &= -\del_i \f_I \,,
&
\d_{J_{i j}} \f_I &= \( x_i \del_j - x_j \del_i \) \f_I \,,
\nn \\
\d_{D} \f_I &= - \( \D_I + x^{\mu} \del_{\mu} \) \f_I \,,
&
\d_{K_i} \f_I &= - 2 x_i \( \D_I + x^{\mu} \del_{\mu} \) \f_I + \( x^\mu x_\mu + A^2 t^2 \) \del_i \f_I + \ldots 
\ee 
where the constant $A$ is a model-dependent function of the $\alpha_I$'s of the background solution.
If $\D_I \neq 0$ (and $\a_I \neq 0$), then the 5 broken generators act non-linearly on the perturbations,
\bea
\nonumber
\d_{P_0} \f_I &=& -\frac{\D_I \alpha_I}{(-t)^{\Delta_I + 1}}  + \ldots \qquad
\d_{J_{0 i}} \f_I = -x_i \frac{\D_I \alpha_I}{(-t)^{\Delta_I + 1}}   + \ldots \\
\d_{K_0} \f_I &=&\( x^\mu x_\mu + A^2 t^2 \) \frac{\D_I \alpha_I}{(-t)^{\Delta_I + 1}}  + \ldots
\label{P0nonlin}
\eea
whereas they act linearly if $\D_I = 0$ (or $\a_I = 0$):
\be
\d_{P_0} \f_I &= - \del_t \f_I
\nc
\d_{J_{0 i}} \f_I = - \( t \del_i + x_i \del_t \) \f_I \,,
\nn \\
\d_{K_0} \f_I &= 2 t \( \D_I + x^{\mu} \del_{\mu} \) \f_I
+  \( x^\mu x_\mu + A^2 t^2 \) \del_t \f_I + \ldots 
\label{P0lin}
\ee
As before, the ellipses indicate terms which do not constrain the quadratic action.

The most general quadratic, two-derivative action for the $\f_I$ which is invariant under spatial translations and spatial rotations is
\be
S = \half \int \dfx \bigg( M^{I J}_1(t) \dot{\f}_I \dot{\f}_J - M^{I J}_2(t) \del_i \f_I \del^i \f_J - M^{I J}_3(t) \f_I \f_J \bigg) ,
\label{genquad1}
\ee
where $M^{I J}_{{\cal I}}$, ${\cal I} = 1,2,3$ are symmetric matrices with arbitrary time dependence. Imposing invariance under dilatations yields the conditions
\be
\dot{M}^{I J}_{1,2} = \frac{2(\D_I - 1)}{t} M^{I J}_{1,2}
\nc
\dot{M}^{I J}_3 = \frac{2(\D_I - 2)}{t} M^{I J}_3 \, . \label{dilcon}
\ee
Since the matrices are symmetric, it follows that $0 = \dot{M}^{I J}_\I - \dot{M}^{J I}_\I = 2(\D_I - \D_J) M^{I J}_\I / t$, hence
\be
(\D_I - \D_J) M^{I J}_\I = 0 \, .
\ee
When $\D_I = \D_J$, the matrix elements $M^{I J}_\I$ are unconstrained, but $M^{I J}_\I = 0$ when $\D_I \neq \D_J$. Thus fields of different conformal weights do not mix at quadratic order in the action. The matrices $M^{I J}_\I$ can therefore be assumed to be block diagonal, with a separate block for each conformal weight.  Within each block,~\eqref{dilcon} implies the time dependence
\be
M^{I J}_{1,2}(t) = m^{I J}_{1,2} \cdot (-t)^{2(\D_I-1)} 
\nc
M^{I J}_3 (t) = m^{I J}_3 \cdot (-t)^{2(\D_I-2)} \, ,
\ee
where the $m^{I J}_\I$'s are constant matrices. Moreover, by redefining the fields, each block of the kinetic matrix can be diagonalized: $m^{I J}_1 \rightarrow \delta_{IJ}$.
Within a block of given conformal weight $\Delta$, the action can therefore be written as
\be
S_\Delta \sim \half \int \dfx (-t)^{2(\D-1)} \( \dot{\f}_I \dot{\f}^I - m^{I J}_2 \del_i \f_I \del^i \f_J - \frac{m^{I J}_3}{t^2} \f_I \f_J \) \,.
\ee

Varying this action with respect to the $\d_{K_i}$ transformation yields
\be
\delta_{K_i} S_\Delta\sim \int \dfx (-t)^{2\D-1}\bigg( \frac{\delta^{IJ}} {\bar{\gamma}^2} - m_2^{IJ}\bigg) \dot{\varphi}_I \partial_j \varphi_J \,,
\ee
where 
\be
\bar{\gamma} \equiv \frac{1}{\sqrt{1- A^2}}\,.
\ee
For this variation to vanish for arbitrary field configurations, the gradient matrix $m_2^{IJ}$ must be proportional to the unit matrix: $m_2^{IJ} = \bar{\gamma}^{-2} \delta^{IJ}$.
The most general action (with a given block of weight $\Delta$) consistent with the linearly realized symmetries is therefore
\be
S_\Delta  \sim \half \int \dfx (-t)^{2(\D -1)} \(  \dot{\f}_I \dot{\f}^I - \frac{1}{\bar{\gamma}^2} \del_i \f_I \del_i \f^I - \frac{m^{I J}_3}{t^2} \f_I \f_J \) \,.
\label{SDelfinal}
\ee

The result is similar to the action~(\ref{quadgen2}) for the ordinary case, except for the sound speed:
at high energy where the mass term can be neglected, perturbations $\varphi_I$ propagate with sound speed
\be
c_s^2 = \frac{1}{\bar{\gamma}^2}\,.
\ee
To avoid gradient instabilities, this sound speed should be real, which requires $A^2 < 1$, in which case $c_s$ is also subluminal. (It is exactly luminal in the case $A = 0$.)

If $\Delta = 0$ or $\alpha_I = 0$, then~(\ref{SDelfinal}) is as far as we can go. Indeed, since the remaining $\d_{P_0}$, $\delta_{J_{0i}}$ and $\delta_{K_0}$ act linearly
on perturbations of this type --- see~(\ref{P0lin}) --- the variation of~(\ref{SDelfinal}) can cancel against the (non-linear) variation of cubic terms in the action involving
one field with $\Delta\neq 0$ and $\alpha_I\neq 0$ (at least one such field must exist to achieve our symmetry breaking pattern). Thus the remaining symmetries impose
no further constraints on the quadratic action alone.

If $\Delta\neq 0$ and the $\alpha_I$ are not all $0$, then the remaining $\d_{P_0}$, $\delta_{J_{0i}}$ and $\delta_{K_0}$ act non-linearly --- see~(\ref{P0nonlin}) --- and therefore constrain
the quadratic action. Invariance under $\d_{P_0}$ implies that the mass matrix within each $\Delta \neq 0$ conformal block obeys the eigenvalue equation:
\be
m^{I J}_3 \a_J = (\D+1) (\D-4)  \a_I \qquad (\Delta\neq 0) \, , \label{timcon}
\ee
where the $\a_I$'s are the coefficients of the background solution~\eqref{phibackgenmany}. This is identical to~(\ref{M3M1p}). The remaining non-linearly realized symmetries $\delta_{K_0}$ and $\delta_{J_{0i}}$ provide no further constraints. 

As a check, we can show that our results are consistent with the quadratic action for the AdS$_5\times S^1$ brane embedding of Sec.~\ref{quadAdS5S1}, consisting of a single field $\varphi$ with $\alpha\neq 0$ and $\Delta = 1$
together with a shift-symmetric $\Delta = 0$ field $\vartheta$. For the $\Delta = 1$ field, the eigenvalue condition~(\ref{timcon}) reduces to $m_3 = -6$, and thus~(\ref{SDelfinal}) becomes
\be
S_\varphi \sim \int \dfx \( \dot{\f}^2 - \frac{1}{\bar{\g}^2}(\del_i \f)^2 + \frac{6}{t^2} \f^2 \)\,.
\ee
For the $\Delta = 0$ field, the assumption of shift symmetry sets $m_3 = 0$, and~(\ref{SDelfinal}) in this case becomes
\be
S_\vartheta\sim \int \dfx t^{-2} \( \dot{\vartheta}^2 - \frac{1}{\bar{\g}^2}(\del_i \vartheta)^2 \) \,.
\ee
These are consistent with~(\ref{quadactiongeom}). 

\section{Coupling to Gravity}
\label{gravity}

We conclude our analysis with a brief discussion of the cosmology that results from the DBI scalar
field theories described above. Irrespective of the details of the theory, it turns out that
the cosmological evolution is completely determined by a single coefficient multiplying the pressure.

As in~\cite{Creminelli:2010ba,Hinterbichler:2011qk,Hinterbichler:2012mv}, we assume that the conformal field theory couples minimally to Einstein gravity, thus mildly breaking conformal
invariance at order $1/M_{\rm Pl}$. Since the background is time-dependent, so are the corresponding energy density
and pressure: $\rho_{\rm CFT} = \rho_{\rm CFT} (t)$ and $P_{\rm CFT} = P_{\rm CFT}(t)$. On the one hand, dilatation invariance implies
that both quantities scale as $1/t^4$. On the other hand, energy conservation implies $\rho_{\rm CFT} = {\rm constant}$ to zeroth order in $1/M_{\rm Pl}$. It follows that
\be
\rho_{\rm CFT} = 0\;,\qquad P_{\rm CFT} = \frac{\beta}{t^4} \,.
\label{rhoP}
\ee
To solve for the resulting cosmological evolution, we must work at next order in $1/M_{\rm Pl}$. Specifically, we can integrate
 the acceleration equation $\dot{H} = - (\rho_{\rm CFT} + P_{\rm CFT})/2M_{\rm Pl}^2$ to obtain the Hubble parameter:
\be
H(t) \simeq \frac{\beta}{6t^3M_{\rm Pl}^2}\,.
\label{Hein}
\ee
As advocated, the cosmological background is fixed in terms of a single parameter $\beta$. In particular, recalling that $t < 0$,
the universe is either slowly contracting for $\beta > 0$, as in the quartic case of~\cite{Hinterbichler:2011qk} or slowly
expanding for $\beta < 0$, as in Galilean Genesis~\cite{Creminelli:2010ba}. 

As an example, let us compute the pressure for the pure DBI theory. This can be done by covariantizing the DBI action~(\ref{dbiaction}), varying
it with respect to the metric, and --- to this order in $1/M_{\rm Pl}$ --- setting the metric to $\eta_{\mu\nu}$. It is easily checked that the energy density vanishes
once~(\ref{galileongammaequation}) is imposed, as it should. Meanwhile, the pressure is
\be
P_{\rm DBI} = \frac{\bar{\gamma}^3}{\bar{\gamma}^2-1} \cdot \frac{1}{t^4} > 0\,,
\ee
where $\bar{\gamma} = 1 + \lambda/4$, from which we can read off $\beta$. The pressure is positive and causes the universe to slowly contract.
In the ``non-relativistic" limit, $\bar{\gamma}\simeq 1$, this reduces to $P_{\rm DBI} \simeq 2/\lambda t^4$, which matches the quartic result of~\cite{Hinterbichler:2011qk}. 

For the DBI Galilean Genesis case, with general action given by~(\ref{galileonaction}), the answer for the pressure depends on the choice of covariantization for $S_3$, {\it i.e.}
whether or not one includes suitable non-minimal couplings for the scalar~\cite{Khoury:2011da}. The brane construction gives a particular prescription for the covariantization~\cite{Goon:2011qf}.
We leave a detailed discussion of the DBI Genesis scenario, including higher-order galileon terms $S_4$ and $S_5$, to future work~\cite{DBIfuture}.

\section{Conclusions}
\label{conclu}

In this paper we have generalized the pseudo-conformal framework to the DBI non-linear realization of the conformal algebra.
As in the original framework, the action for perturbations is fixed, up to a few parameters, by the symmetry breaking pattern.
An important difference with the original framework is a universal and strictly {\it subluminal} speed of propagation for perturbations. 

The pure DBI version of the scenario, discussed in Sec.~\ref{dbisymmetrybreaking}, is a ``relativistic" extension of the quartic
model of Rubakov~\cite{Rubakov:2009np}. The upshot is a geometric interpretation of the scenario in terms of a brane moving in an AdS$_5$ bulk space-time.
The weight-0 fields required to generate scale invariant perturbations also have a natural geometric origin as isometries along
additional extra dimensions, the simplest example of which is the AdS$_5 \times S^1$ geometry studied in Sec.~\ref{dbiscaleinvariance}.
This opens the door to the search for UV completions of the scenario through explicit realizations in string theory compactifications, analogous
to brane inflation constructions. From a phenomenological perspective, it would be interesting to generalize the coset derivation of the effective action to the DBI non-linear realization,
as well as to derive the associated Ward identities~\cite{paolomarkoaustin}.

We also derived a DBI version of the Galilean Genesis scenario by including the cubic DBI conformal galileon term.
As in the original Genesis scenario, stability around the $1/t$ solution requires that the theory be unstable around a trivial background,
like in the ghost condensate~\cite{ArkaniHamed:2003uy}. In the DBI version, this corresponds to the requirement that the brane have negative tension.
The upshot of the DBI realization is a subluminal sound speed for perturbations, which therefore offers a cure for the
superluminality issues of the original Genesis scenario. In forthcoming work~\cite{DBIfuture}, we will study the most general version of DBI Galilean Genesis, 
including all 5 DBI conformal galileon terms. Once again this should lead to stable, Null-Energy violating $1/t$ solutions, with strictly
subluminal propagation, but it will be interesting to see if these theories can at the same time be stable around trivial backgrounds, corresponding
to positive-tension branes. \\
\\
{\bf Acknowledgments:} We thank Paolo Creminelli, Alberto Nicolis and Enrico Trincherini for helpful discussions. This work is supported in part by
the US DOE (G.E.J.M.), NASA ATP grant NNX11AI95G (A.J.), the Alfred P. Sloan Foundation
and NSF CAREER Award PHY-1145525 (J.K.), as well as  funds from the University of Pennsylvania (K.H.). Research at the Perimeter Institute is supported by the Government of Canada through Industry Canada and by the Province of Ontario through the Ministry of Research and Innovation. 

\appendix
\section{Relation Between Parameterizations of the Algebra}
\label{cosetrelations}
Throughout the text, we have focused on starting with the nonlinear parameterization of the conformal algebra where the field in the unbroken phase transforms nonlinearly under conformal transformations
\be
\delta_{D}\tilde\pi=-1 - x^\nu \del_\nu\tilde\pi~,~~~~~~~~~~~
\delta_{K_\mu}\tilde\pi= -2 x_\mu + \left(e^{-2\tilde\pi}+x^2\right)\partial_\mu\tilde\pi-2x_\mu x^\nu\partial_\nu\tilde\pi~.
\ee
The Lagrangians invariant under these symmetries which have second order equations are the {\it conformal DBI galileon} terms \cite{deRham:2010eu,Goon:2011qf}
\begin{align}
\nonumber
{\cal L}_1 =&~e^{4\tilde\pi}\,;\\ \label{firstfewlag}
{\cal L}_2 =&~e^{4\tilde\pi}\sqrt{1+e^{-2\tilde\pi}(\partial\tilde\pi)^2}\,;\\\nonumber
{\cal L}_3 =&~-\g^2\partial_\mu\tilde\pi\partial^\mu\partial^\nu\tilde\pi\partial_\nu\tilde\pi+e^{-2\tilde\pi}\square\tilde\pi+e^{4\tilde\pi}(\g^2-5)~,\\\nonumber
&\vdots
\end{align}
where $\gamma^{-1} \equiv \sqrt{1+e^{-2\tilde\pi}(\partial\tilde\pi)^2}$, and where we have made a field redefinition $e^{\tilde\pi}=\phi$ with respect to the rest of the text. Here we comment on the relation between this situation and that considered in \cite{Creminelli:2010ba}, where the field in the unbroken phase, $\pi$, {\it also} transforms non-linearly under SCTs and dilations
\be
\delta_D\pi = -1-x^\mu\partial_\mu\pi~,~~~~~~~~~~~\delta_{K_\mu}\pi =-2x_\mu -(2x_\mu x^\nu\partial_\nu-x^2\partial_\mu)\pi~.
\ee
Here the first few Lagrangians invariant under these symmetries and possessing second order equations are the {\it conformal galileon} terms
\begin{align}
\nonumber
{\cal L}_1 =&~e^{4\pi}\,;\\
{\cal L}_2 =&~-\frac{1}{2}e^{2\pi}(\partial\pi)^2\,;\\\nonumber
{\cal L}_3 =&~2\square\pi(\partial\pi)^2+(\partial\pi)^4~,\\\nonumber
&\vdots
\end{align}
Although the symmetry algebras and breaking patterns are the same in both theories ---  they both non-linearly realize conformal symmetry and linearly realize Poincar\'e symmetry --- the theories appear to be physically inequivalent in that perturbations around the $1/t$ solution in the conformal galileon theory propagate exactly luminally, while perturbations around the $1/t$ solution propagate at less than the speed of light.

This is slightly disconcerting because both of these theories may be constructed as the theory of the Goldstone of spontaneously broken conformal symmetry via the well known coset construction \cite{Coleman:1969sm,Callan:1969sn, volkov}. The uniqueness of the coset construction has been established for internal symmetries, but we know of no such proof for the case when space-time symmetries are broken. 
There are two possibilities: either there is a field redefinition transforming one theory into the other which preserves the background profile, or the coset construction does not guarantee a unique low-energy Lagrangian in the space-time symmetry case.

In~\cite{Bellucci:2002ji}, an explicit map was constructed between the two realizations using coset construction machinery.  This map preserves the $1/t$ backgrounds, so it should be the equivalence we are asking for.   However, it is inherently non-local, since under the field redefinition the coordinates on one side get mixed with fields on the other side.\footnote{The fact that fields and coordinates get mixed in a non-trivial way can be understood simply by noting that, in the notation of~\cite{Bellucci:2002ji}, the set $(x^\mu, \phi, \Omega_\mu)$ parameterizes the same coset space as the set $(y^\mu, \pi, \xi_\mu)$, where $\Omega_\mu$ and $\xi_\mu$ are additional fields necessary to parameterize the full coset but which are non-dynamical. Then, changing coordinates from one set to the other will generically mix the fields and coordinates in a non-trivial way.} This may also be seen through the fact that operators on one side get mapped to an infinite series of operators on the other side.

Even if these theories are equivalent by themselves, they will surely be different once we couple to matter and gravity, due to the fact that the field redefinition that relates them mixes fields and coordinates. Even if we minimally couple the conformal DBI theory to gravity, after mapping to the conformal galileon theory, there will be non-minimal couplings. Therefore, it is worthwhile to study the theory in these variables, even if without gravity it turns out that the theory is actually equivalent to the theory of the conformal galileons.

\end{document}